\documentclass[12pt,psfig,a4]{article}
\usepackage{float}
\input psfig.sty
\begin{document}
\topmargin -0.5cm
\textheight=21cm
\evensidemargin=1.2cm
\newcommand{\salut}[1]{\centerline{\parbox{13.5cm}{\caption{\small #1}}}}
\thispagestyle{empty}
\noindent\hbox to\hsize{April 2000 \hfill  ULG-PNT-00-1-IR}\\
\vskip 1.6in
\begin{center}
{\large \bf Helicity in diffractive vector-meson production.
}\\
\vskip .3in
I. Royen\footnote{royen@nuclth02.phys.ulg.ac.be}\\{\small
Inst. de Physique, U. de Li\`ege,
B\^at. B-5, Sart Tilman, B4000 Li\`ege, Belgium}\\
\vskip 1in
{\bf Abstract}
\end{center}
\begin{quote}
We study the helicity amplitudes describing the quasielastic production of vector mesons in deep inelastic scattering within the context of the model which we previously introduced to describe the ratio of longitudinal to transverse cross sections.  We calculate here a full set of spin flip (and non-flip) amplitudes and naturally find a significant violation of s-channel helicity conservation.
We present predictions for the 15 spin-density matrix elements which completely define the angular distributions and the helicity properties of the produced meson.
\end{quote}

\newpage
\section{Introduction}

\noindent
Elastic vector meson production in photon-proton scattering $\gamma^* p \rightarrow Vp$ is an important
process under intensive experimental and theoretical study. It should provide us with information on the quark and gluon structure of hadrons as well as information on the exchange forces between the particles in this process.
\\

\noindent
From many papers (see references in \cite{Crittenden}), it turns out that perturbative QCD models where pomeron is represented by two-gluon exchange are able to reproduce the main features of the HERA data.
In a previous paper \cite{IR2}, with J.R. Cudell, we implemented Fermi momentum in elastic vector-meson production and proposed a new approach which allows the quarks to be off-shell, and which naturally reproduces the data.
The $Q^2$ and $m_V$ dependence of the dominant transitions $\gamma^*_L \to V_L$ and $\gamma^*_T \to V_T$ are in good agreement with the data and the model naturally reproduces the ratio $\sigma_L/\sigma_T$ (L and T stand for the longitudinal and transverse polarisations).  The plateau observed experimentally comes from the interplay between contributions from on-shell and off-shell quarks, which have different asymptotic behaviours.
\\

\noindent
The aim of this paper is to explore the polarisation effects in quasielastic electroproduction of vector mesons within the framework of the above model.  
We know that the cross section for the exclusive production of vector mesons from virtual photon has contributions from both transverse and longitudinal photons.  What about the spin of the produced meson ?
Experimentally, information about the polarisation state of the produced meson is extracted from the angular distributions of the meson decay products ($\pi^+ \pi^-$ for the $\rho$ meson).
Previous studies at HERA \cite{HERA1,HERA2} were consistent with s-channel helicity conservation (SCHC), i.e. the produced meson retains the helicity of the incoming virtual photon.
Sufficient data are now available to test the validity of SCHC at HERA by measuring the full set of matrix elements which completely determine the angular distributions of the decay.  As a consequence, H1 \cite{Barbara} and ZEUS \cite{helZEUS} have found a small but significant violation of SCHC in $\rho$ meson production.\\

\noindent
Following the formalism introduced by Schillings \cite{shillings}, the 15 spin density matrix elements, $r_{ij}^{\alpha \beta}$, which completely define the angular distributions $W(\rho \rightarrow \pi^+ \pi^-)$, are related to various combinations of the helicity amplitudes $A_{\lambda_V \lambda_N,\lambda_\gamma \lambda_{N'}}$, where $\lambda_V$ and $\lambda_\gamma$ are, respectively, the helicities of the vector meson and of the photon, and $\lambda_N$ and $\lambda_{N'}$ those of the incoming and outgoing proton (see appendix).
\\

\noindent
To get a feeling for these amplitudes we can consider some special cases, as well as general constraints.
 
- Under parity conservation in the $t$-channel, 
the helicity amplitudes yield the following symmetry relation:
\begin{equation}
A_{-\lambda_V \lambda_{p'}, -\lambda_\gamma \lambda_p}=(-1)^{\lambda_V - \lambda_\gamma} A_{\lambda_V \lambda_{p'}, \lambda_\gamma \lambda_p}.
\label{eq:parite}
\end{equation}
There are then five independent helicity amplitudes: the two helicity conserving amplitudes, two single spin-flip amplitudes and one helicity double-flip amplitude\footnote{
We shall discuss reactions with unpolarised protons, therefore the proton can be formally considered as a spinless target and we shall indicate only the polarisation states of the virtual photon and the produced meson.  L and T stand then for the longitudinal and transverse polarisations.}:
\begin{eqnarray}
A_{V_L \gamma^*_L}&=&A_{00},\\
A_{V_T \gamma^*_T}&=&A_{11}=A_{-1-1},\\
A_{V_L \gamma^*_T}&=&A_{01}\ {\rm with}\ A_{0\ -1}=-A_{01},\\
A_{V_T \gamma^*_L}&=&A_{10}\ {\rm with}\ A_{-1\ 0}=-A_{10},
\end{eqnarray}
\begin{equation}
{\rm and}\ \ A_{-1 1}=A_{1 -1}
\end{equation}

- Under the assumption of $s$-channel helicity conservation (SCHC), 
the helicity of the virtual photon is retained by the vector meson V:
\begin{equation}
A_{\lambda_V \lambda_{p'}, \lambda_\gamma \lambda_p}= A_{\lambda_V \lambda_{p'}, \lambda_\gamma \lambda_p}\ \delta_{\lambda_V \lambda_\gamma}\ \delta_{\lambda_{p'} \lambda_p}.
\end{equation}
There are only two independent helicity amplitudes,
single and double helicity flip amplitudes are then zero:
\begin{eqnarray}
A_{\lambda_V \lambda_\gamma}=0,\ \ \ \ \ \ \ \lambda_V \not= \lambda_\gamma
\end{eqnarray}
The assessment of the validity of this last assumption is the object of this letter.\\

\noindent
Theoretical studies of helicity amplitudes for diffractive production of vector meson at large $Q^2$ have first been performed by Ivanov and Kirschner \cite{Ivanov}, and later by Nikolaev et al \cite{Nikolaev} using perturbative QCD, and extending the QCD factorisation theorem.  Although there is no longer any doubt about the dominance of transitions $\gamma^*_L \rightarrow V_L$ and $\gamma^*_T \rightarrow V_T$, they reported a substantial s-channel helicity non conservation.  They assumed that all helicity amplitudes (except the double-flip) are proportional to the gluon structure function of the proton and they agree that the largest amplitude violating SCHC is $A_{10}$, where a transverse photon produce a longitudinal vector meson.\\

\noindent
This study differs from the previous ones as now all helicity properties are coming from the transition $\gamma^* \rightarrow V$ only.
Hence in this paper, we concentrate only on the upper diagram (Fig.~\ref{fig:feynmanTL}), following our previous model \cite{IR2} to calculate all helicity amplitudes, we extend our results to lower $Q^2$ and we derive the 15 matrix elements to be compared with the data. 
We first present the main steps of the model described in \cite{IR2} and calculate the helicity amplitudes.  Their properties are given in section 3.  In section 4, we calculate the matrix elements to compare with the data.  We summarise our conclusions in section 5.\\

\newpage
\section{Kinematics and calculation of the different helicity amplitudes.}
\begin{figure}[H]
\centerline{
\psfig{figure=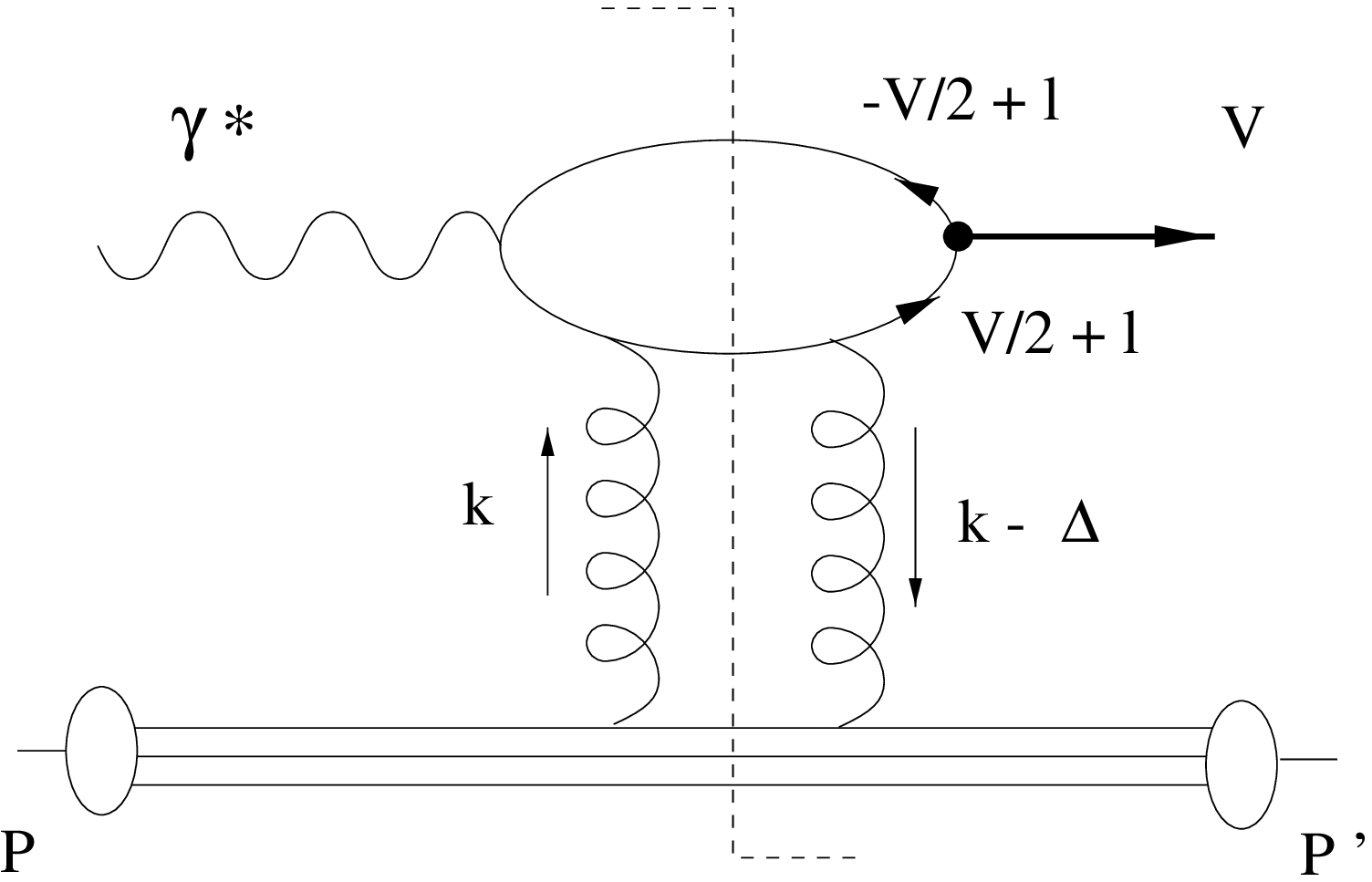,height=4cm}\ \ \ \ \
\psfig{figure=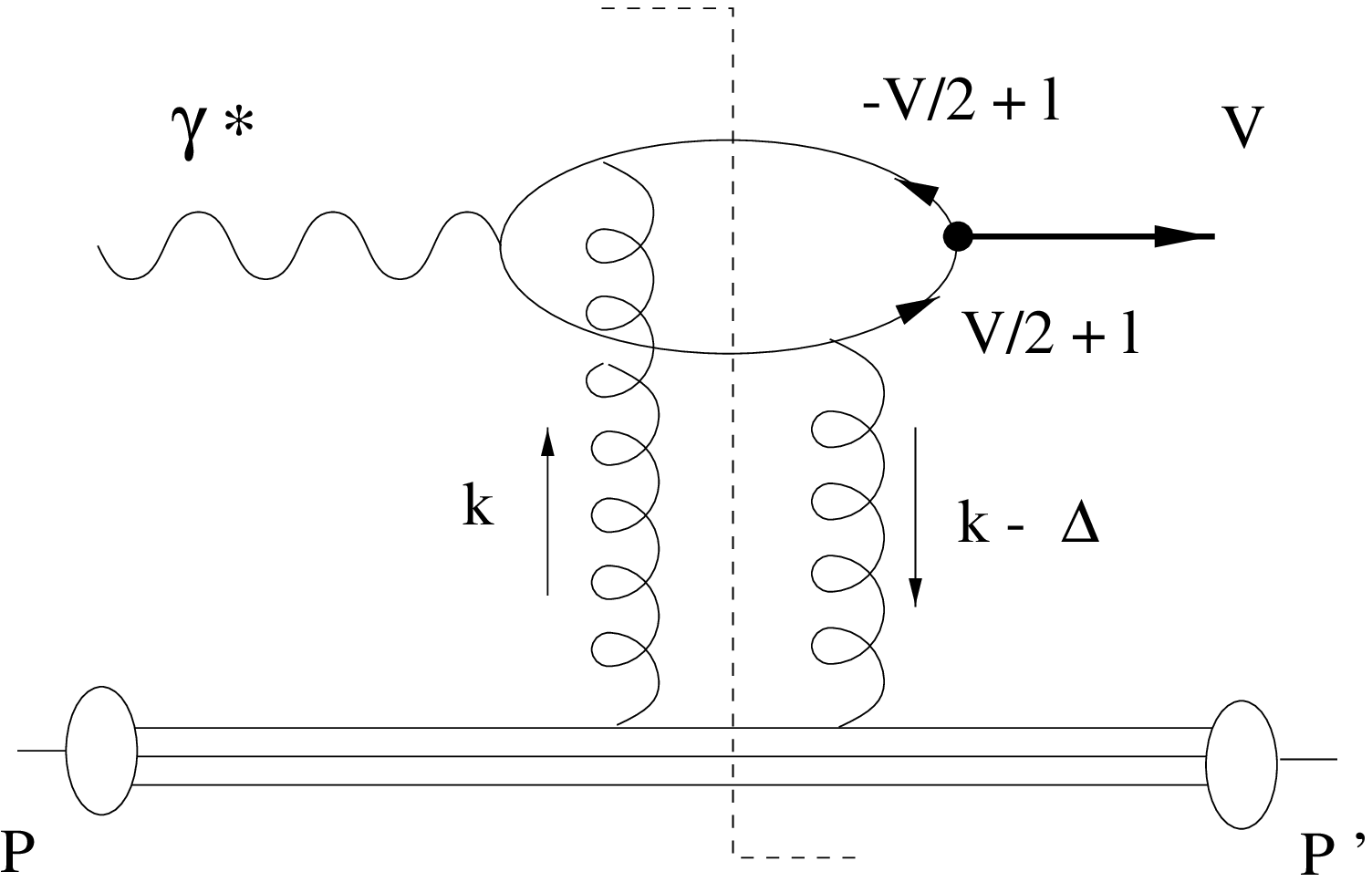,height=4cm}}
\salut{The two diagrams accounting for the transition $\gamma p \rightarrow V p$. The dashed line represents the cut which puts the intermediate state on-shell.\label{fig:feynmanTL}}
\end{figure}

One usually assumes that the exclusive process at high energy proceeds from the emission of a pair of gluons from the proton, which interact with a $q\bar{q}$ pair emerging from the photon, and which transfer momentum so that this pair can be turned into a vector meson, as shown in Fig.~\ref{fig:feynmanTL}.
The following kinematic variables are used to describe the problem $\gamma^* p \rightarrow Vp$ \cite{IR2}.  The photon has 4-momentum $q$ and polarisation $\epsilon(\lambda_\gamma)$, with $q.q=-Q^2$.  We define
\begin{equation}
\epsilon(\lambda_\gamma=\pm1)=\mp (1/\sqrt{2})(0,1,\pm i,0)
\end{equation} 
which correspond to circularly polarised radiation for transverse photons and 
\begin{equation}
\epsilon(\lambda_\gamma=0)=(1/\sqrt{Q2})(q^3,0,0,q^0)
\end{equation} 
for longitudinal photons.  
The hard process generating the meson, and the $q\bar{q} \rightarrow V$ amplitude are treated together through the introduction of a meson vertex function.  The gluons will couple to the quarks emerging from this vertex. 
The vector meson has momentum $V=q+\Delta$ and polarisation e($\lambda_V$) defined in the same way as $\epsilon(\lambda_\gamma)$.  With $t=\Delta^2$, $V^2=m_V^2$, hence $\Delta.q=(m_V^2-t+Q^2)/2$.  The quarks composing the meson are written as $v+l$ and $-v+l$ with $v=V/2$.
\\

\noindent

\noindent
We assume the vertex function to be described by:
\begin{eqnarray}
\Gamma_\mu&=&\Phi(l)\gamma_\mu\\
{\rm with}\ \Phi(l)&=&N e^{-{bf L}^2 \over 2 p_F^2}
\end{eqnarray}
where ${\bf L}^2$ is the quark 3-momentum in the meson rest frame, and where $p_F$ is a Fermi momentum scale equal to $0.3$ GeV in the $\rho$ and $\phi$ cases.
As we do not want to look in detail at the scattered proton, we only consider its three valence quarks of momentum $p$, we assume that we can neglect its mass and we put it on-shell $p.p=0$.
Moreover, we introduce the following form factor ${\cal F}(k, \Delta)=3({\cal E}_1(t)-{\cal E}_2(k,k-\Delta))$ where:
\begin{equation}
{\cal E}_1(t=\Delta^2)\simeq {(3.53 -2.79t) \over (3.53-t)(1-t/0.71)^2}
\label{infrarouge1}
\end{equation}
is the quark-level form factor when both gluons hit the same quark line, and
\begin{equation}
{\cal E}_2(k,k-\Delta)={\cal E}_1(k^2+(k-\Delta)^2-k.(k-\Delta))
\label{infrarouge2}
\end{equation}
if the gluons hit different quark lines.  These form factors, which take the dipolar character of the proton into account, are necessary to obtain IR finiteness.\\

\noindent
This model allows us to calculate all the helicity amplitudes with and without spin flip between the photon and the vector meson.
We shall be working in the high-$w^2$ limit, and we write $p.q=(p+q)^2/2\approx w^2/2$.
As we expect the amplitude to be $w$-independent, we shall be calculating the discontinuity of the amplitude using Cutkovsky's rules and putting the intermediate quark propagators on-shell.
The transition amplitude is given by the convolution between the upper diagrams and the lower diagrams (Fig.~\ref{fig:feynmanTL}):
\begin{eqnarray}
A_{\lambda_V \lambda_\gamma}&=& {2 \over 3}\ (4 \pi \alpha_S)^2 g^{elm} e_Q\ w^2\nonumber\\
&\times&\ \int\ {d^4 l \over (2 \pi)^4}\ {d^4k \over (2\pi)^4}\ {\cal F}(k, \Delta)\ {1 \over \sqrt{3}} \Phi(l)\nonumber\\
&\times&\ {4(p_\alpha p_\beta)T^{\alpha \beta}_{\lambda_V \lambda_\gamma}\over k^2 (k-\Delta)^2}.
\label{eq:T_amplitude}
\end{eqnarray}
where $4(p_\alpha p_\beta)$ is the leading contribution of the lower quark lines and $T^{\alpha \beta}$ the sum of the two cut diagrams:
\begin{eqnarray}
T^{\alpha \beta}&=&{\big [}{T^{\alpha \beta}_1(\epsilon,e)\over (q-v+l).(q-v+l)-m_q^2}+{T^{\alpha \beta}_2(\epsilon,e)\over (-v+l+k).(-v+l+k)-m_q^2}{\big ]}\nonumber\\
&\times&{((2\pi)^2 \delta((-v+l)^2-m_q^2)\delta((v+l+k-\Delta)^2-m_q^2)\over (v+l)^2-m_q^2}.
\end{eqnarray}

\noindent
Following \cite{IR2}, the traces of the upper bubbles of the graphs are described by:
\begin{eqnarray}
T^{\alpha \beta}_1&=&Tr\{\gamma.e[\gamma.(v+l)-m_q]\gamma_\beta[\gamma.(q-v+l+k)+m_q]\gamma^\alpha\nonumber\\
&\times&[\gamma.(q-v+l)+m_q]\gamma.\epsilon[\gamma.(-v+l)+m_q]\}\\
T^{\alpha \beta}_2&=&Tr\{\gamma.e[\gamma.(v-l)-m_q]\gamma_\alpha[\gamma.(v-l-k)+m_q]\gamma^\epsilon\nonumber\\
&\times&[\gamma.(v-q-k-l)+m_q]\gamma.\beta[\gamma.(-v-l)+m_q]\}
\end{eqnarray}

\noindent
Using equation (\ref{eq:T_amplitude}), we calculate:
\begin{itemize}
\item the longitudinal vector meson production amplitude by a longitudinal photon: $T_{\lambda_V \lambda\gamma}=T_{00}$,
\item the transverse-vector-meson-production amplitude by a transverse photon: 
\begin{itemize}
\item with single-spin flip: $T_{\lambda_V \lambda\gamma}=T_{11}$, $T_{-1-1}$,
\item with double-spin flip: $T_{1-1},T_{-11}$, 
\end{itemize}
\end{itemize}
\begin{itemize}
\item single-spin-flip helicity amplitudes where: 
\begin{itemize}
\item a transverse photon produces a longitudinal meson:\\
\indent \hskip 2cm $T_{\lambda_V \lambda\gamma}=T_{01},T_{0-1}$,
\item a longitudinal photon produces a transverse meson:\\
\indent \hskip 2cm $T_{\lambda_V \lambda\gamma}=T_{10},T_{-10}$.
\end{itemize}
\end{itemize}

\subsection{SCHC amplitudes.}
The results obtained with this model confirm the dominance of the longitudinal amplitude $T_{00}$ at high $Q^2$,
in agreement with \cite{Brodsky2,Martin,Rysk}.
In the high-energy $w$ limit, defining $l={\alpha \over 2w^2}p\ + {\beta \over 2}q\ +{{\bf l}_t \over 2}$ and $k={\zeta \over 2w^2}p\ + {\xi \over 2w^2}q\ + {{\bf l}_t \over 2}$, the longitudinal-vector-meson-production amplitude by a longitudinal photon $T_{00}$ is the following:
\begin{eqnarray}
T_{00}&=&{-4\ (\mu_q^2+(1-\beta^2) m_V^2 -{\bf l}_t^2)\ (1+\beta)^2\ (1-\beta)\ Q\ \epsilon_L.e_L \over [(1-\beta)^2\ t\ -\mu_q^2 +{\bf l}_t^2 -(1-\beta^2)Q^2 -2(1-\beta)\ {\bf l}_t.\Delta_t]\ m_V}\nonumber\\
&\times&{2(1-\beta) {\bf k}_t.\Delta_t -2{\bf k}_t.{\bf l}_t +{\bf k}_t^2 \over D(Q^2,\ t,\ k_t,\ l_t)}
\label{eq:ALL}
\end{eqnarray}
where
\begin{eqnarray}
D(Q^2,\ t,\ k_t,\ l_t)&\equiv& [\ (1-\beta^2)Q^2 -(1-\beta)^2\ t+\mu_q^2\ -{\bf l}_t^2\ +{\bf k}_t^2 \nonumber\\
&&\ \ +2(1-\beta){\bf k}_t.\Delta_t\ -2{\bf k}_t.{\bf l}_t\ +2(1-\beta){\bf l}_t.\Delta_t\ ].
\end{eqnarray}
and $\mu_q=2m_q$ the mass of the quarks in the upper loop of the diagram.\\

\noindent
The transverse amplitude $T_{11}$ ($\lambda_V=\lambda_\gamma=+1$), is more complicated.  To give an analytical expression, we shall concentrate on the case $t=0$ \cite{IR2}:
\begin{eqnarray}
&&T_{11}(t=0)={-8\ (1+\beta)\over [(1-\beta^2)Q^2\ +\mu_q^2 -{\bf l}_t^2\ +{\bf k}_t^2 -2{\bf l}_t.{\bf k}_t][(1-\beta^2)Q^2\ +\mu_q^2 -{\bf l}_t^2]}\nonumber\\
&\times&{\Large \{}\ [{\bf l}_t^2\ -(1-\beta^2) Q^2\ -\mu_q^2]\ (\epsilon_t.{\bf l}_t\ e_t.{\bf k}_t\ -\beta^2\ \epsilon_t.{\bf k}_t\ e_t.{\bf l}_t\ - {\bf l}_t.{\bf k}_t\ \epsilon_t.e_t)\nonumber\\
&&\ \ +(1-\beta^2)\ ({\bf k}_t^2 -2{\bf l}_t.{\bf k}_t)\ \epsilon_t.{\bf l}_t\ e_t.{\bf l}_t\ +({\bf l}_t^2- \mu_q^2)(2 {\bf l}_t.{\bf k}_t\ +{\bf k}_t^2)\ \epsilon_t.e_t\ {\Large \}}\nonumber\\
\label{eq:ATT}
\end{eqnarray}

\subsection{The spin-flip amplitudes.}
The calculation are similar than for the SCHC amplitudes.
As the amplitudes conserve the parity at the vertex
\begin{eqnarray}
{\cal A}_{-\lambda_V, -\lambda_\gamma}&=&(-1)^{\lambda_V - \lambda_\gamma} {\cal A}_{\lambda_V, \lambda_\gamma}\nonumber
\end{eqnarray}
we have only three spin-flip amplitudes to calculate (${\cal A}_{01}$,${\cal A}_{10}$,${\cal A}_{-11}$).
\\

\noindent
The two amplitudes ${\cal A}_{01}$ et ${\cal A}_{10}$, derived from the Feynman diagrams (fig. \ref{fig:feynmanTL}), are given by:
\begin{eqnarray}
T_{01}(\lambda_V=0,\lambda_\gamma=+1)&=& { N_{01} \over D},
\label{eq:heleq1}\\
T_{10}(\lambda_V=+1,\lambda_\gamma=0)&=& { N_{10} \over D},
\label{eq:heleq2}
\label{eq:ATLeq}
\end{eqnarray}
with :
\begin{eqnarray}
D&=& [\ 2(1-\beta)\ {\bf l}_t.\Delta_t\ -{\bf l}_t^2\ +\mu_q^2\ +(1-\beta^2) Q^2\ -(1-\beta)^2\ t\ ]\nonumber\\
&\times&[\ 2(1-\beta)\ {\bf l}_t.\Delta_t\ + 2(1-\beta)\ {\bf k}_t.\Delta_t\ -2 {\bf l}_t.{\bf k}_t\nonumber\\
&&\ \ \ \ \ \ \ \ \ + {\bf k}_t^2\ -{\bf l}_t^2\ + \mu_q^2\ +(1-\beta^2)Q^2\ -(1-\beta)^2\ t\ ],\\
\nonumber\\
N_{01}&=&{4 \over m_V}\ \beta\ (1+\beta)\ [\ {\bf l}_t^2 -\mu_q^2 -(1-\beta^2)m_V^2\ ]\nonumber\\
&\times&{\Large \{}\ \epsilon_t.\Delta_t\ (1-\beta)\ [4(1-\beta)\ \Delta_t.{\bf k}_t\ +2{\bf l}_t.{\bf k}_t\ +{\bf k}_t^2]\nonumber\\
&+&\ \ \ 2\epsilon_t.{\bf l}_t\ [{\bf l}_t.{\bf k}_t -(1-\beta) \Delta_t.{\bf k}_t]\nonumber\\
&+&\ \ \ \epsilon_t.{\bf k}_t\ [2(1-\beta) \Delta_t.{\bf l}_t\ -{\bf k}_t^2\ -{\bf l}_t^2\ +\mu_q^2\ +(1-\beta^2) Q^2\ -(1-\beta)^2\ t\ ]\ {\Large \}},\nonumber\\
\\
N_{10}&=&-8\ \beta\ \epsilon_L.q\ (1+\beta)^2 (1-\beta)\ {\bf l}_t.e_t\ [\ 2 {\bf k}_t.{\bf l}_t\ - 2(1-\beta) {\bf k}_t.\Delta_t\ -{\bf k}_t^2\ ].\label{reff}
\end{eqnarray}

\noindent
Performing the numerical calculations, we observe the following hierarchy:
\begin{equation}
|{\cal A}_{00}| >|{\cal A}_{11}| >|{\cal A}_{01}| >|{\cal A}_{10}|>...
\end{equation}
for relatively high $Q^2$.
In the HERA kinematical range \cite{Barbara,helZEUS}, we find that the helicity amplitude $|{\cal A}_{10}|$, where a transverse meson is produced by a longitudinal photon, is of the order of 20 times smaller than the amplitude for the production of a longitudinal meson by a transverse photon $|{\cal A}_{01}|$, itself 30 times smaller than the longitudinal amplitude ${\cal A}_{00}$.
The double spin-flip amplitudes $A_{1-1}=A_{-11}$ are still smaller and are neglected in this work.\\

\section{Amplitudes properties.}
\subsection{Case of a zero Fermi momentum: $l=0 \rightarrow \beta = 0$.}
The single spin-flip amplitudes ${\cal A}_{01}$ (\ref{eq:heleq1}) and ${\cal A}_{10}$ (\ref{eq:heleq2}) are proportional to the Sudakov coefficient $\beta$ of the quark Fermi momentum ($l={l_t \over 2}+{\beta \over 2}q+ {\alpha \over 2 w^2}p$).  
When $\beta=0$ these ones become zero at the opposite of the two helicity conserving amplitudes ${\cal A}_{00}$ (\ref{eq:ALL}) et ${\cal A}_{11}$ (\ref{eq:ATT}).  As we can see, the Fermi momentum plays an important role in the presence of ${\cal A}_{01}$ and ${\cal A}_{10}$.
The conditions $\beta =0$ or $l_{\rm Fermi}=0$ imply $s$-channel helicity conservation.

\subsection{$Q^2$ behaviour.}
\begin{figure}[H]
\centerline{
\psfig{figure=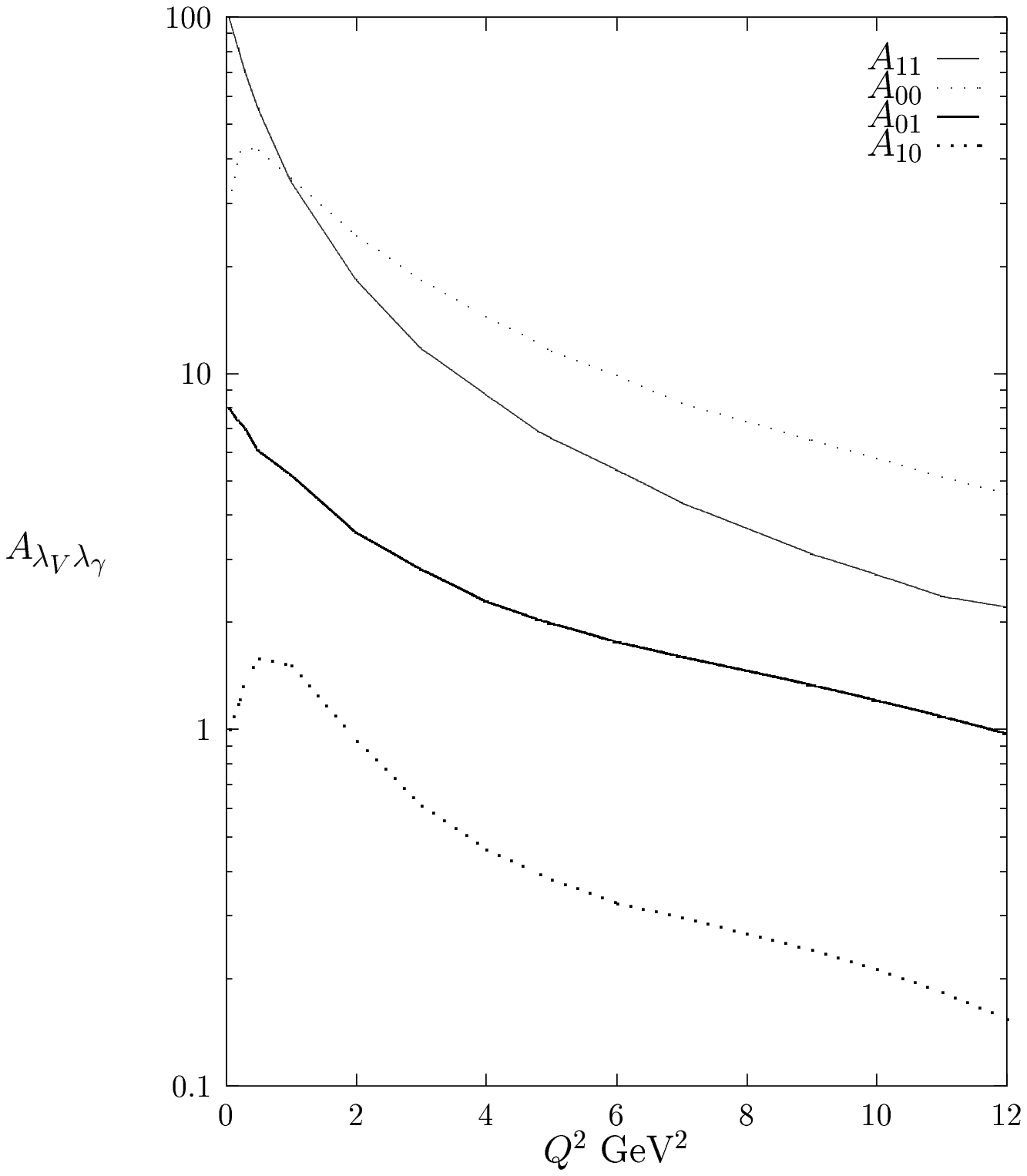,height=10cm}}
\salut{$Q^2$ dependence of the helicity amplitude for $\langle |t| \rangle=0.1352$ GeV$^2$. \label{fig:T_Q2}} 
\end{figure}

Figure \ref{fig:T_Q2} shows the $Q^2$ evolution of the different helicity amplitudes for a mean value of $|t|$ fixed ($|t|=0.1352$ GeV$^2$).  We clearly see that the vector meson production amplitudes by a real longitudinal photon ($Q^2=0$), $A_{00}$ and $A_{10}$, decrease to zero when $Q^2 \rightarrow 0$.
\\

\noindent
In photoproduction, as expected from the Ward identity, the expressions (\ref{eq:ALL}) and (\ref{reff}) are exactly zero:
\begin{eqnarray}
T_{00}&=&0,\nonumber\\
T_{10}&=&0,\nonumber\\
\end{eqnarray}
whereas $T_{11}\not= 0$ and $T_{01}\not= 0$.\\
\noindent
For high-$Q^2$, looking at the numerical results (fig.~\ref{fig:T_Q2}), we observe that the behaviour is more complex than expected by just taking the asymptotic limit of equations (\ref{eq:ALL}, \ref{eq:ATT}, \ref{eq:heleq1} and \ref{eq:heleq2}).
The integration over $l^2$ and $\beta$ introduces supplementary $Q^2$ terms in the contribution to $T_{11}$, $T_{01}$ and $T_{10}$ due to the off-shell quark.
The $Q^2$ range at HERA is far from the asymptotic limit, as consequence, we are enable to control the asymptotic calculation of the amplitudes.

\subsection{Low ${\bf k}_t^2$ behaviour.}
As in the previous paper \cite{IR2}, the amplitudes are finite in the infrared region.  Hence, (\ref{eq:T_amplitude}) gives:
\begin{equation}
d{\cal A}_{\lambda_V \lambda_\gamma} \propto {T_{\lambda_V \lambda_\gamma}\ {\cal F}(k,k-\Delta) \over {\bf k}_t^2 ({\bf k}_t-\Delta_t)^2}.
\end{equation}

\noindent
If we develop the $T_{\lambda_V \lambda_\gamma}$ expressions (eqs.\ref{eq:ALL}, \ref{eq:ATT}, \ref{eq:heleq1} and \ref{eq:heleq2}) in the region of small transverse gluon momentum ${\bf k}_t^2\to 0$, we obtain:
\begin{equation}
d{\cal A}_{\lambda_V \lambda_\gamma}({\bf k}_t^2=0)\propto{{\cal F}(k,k-\Delta)\over ({\bf k}_t -\Delta_t)^2}
\end{equation}
As the proton form factor can be expressed by ${\cal F}(k,k-\Delta) \propto {\bf k}_t.({\bf k}_t-\Delta)$ in the infrared region \cite{IR1}, all subsistent singularity of the denominator is canceled.  

\subsection{ $|t|=0$ behaviour.}
As $|t|=|t_{min}| \approx 0$, we observe the cancellation of all the spin-flip amplitudes:
\begin{equation}
{\cal A}_{01}={\cal A}_{0-1}={\cal A}_{10}={\cal A}_{-10}={\cal A}_{1-1}={\cal A}_{-11}=0.
\end{equation}

\begin{figure}[H]
\centerline{
\psfig{figure=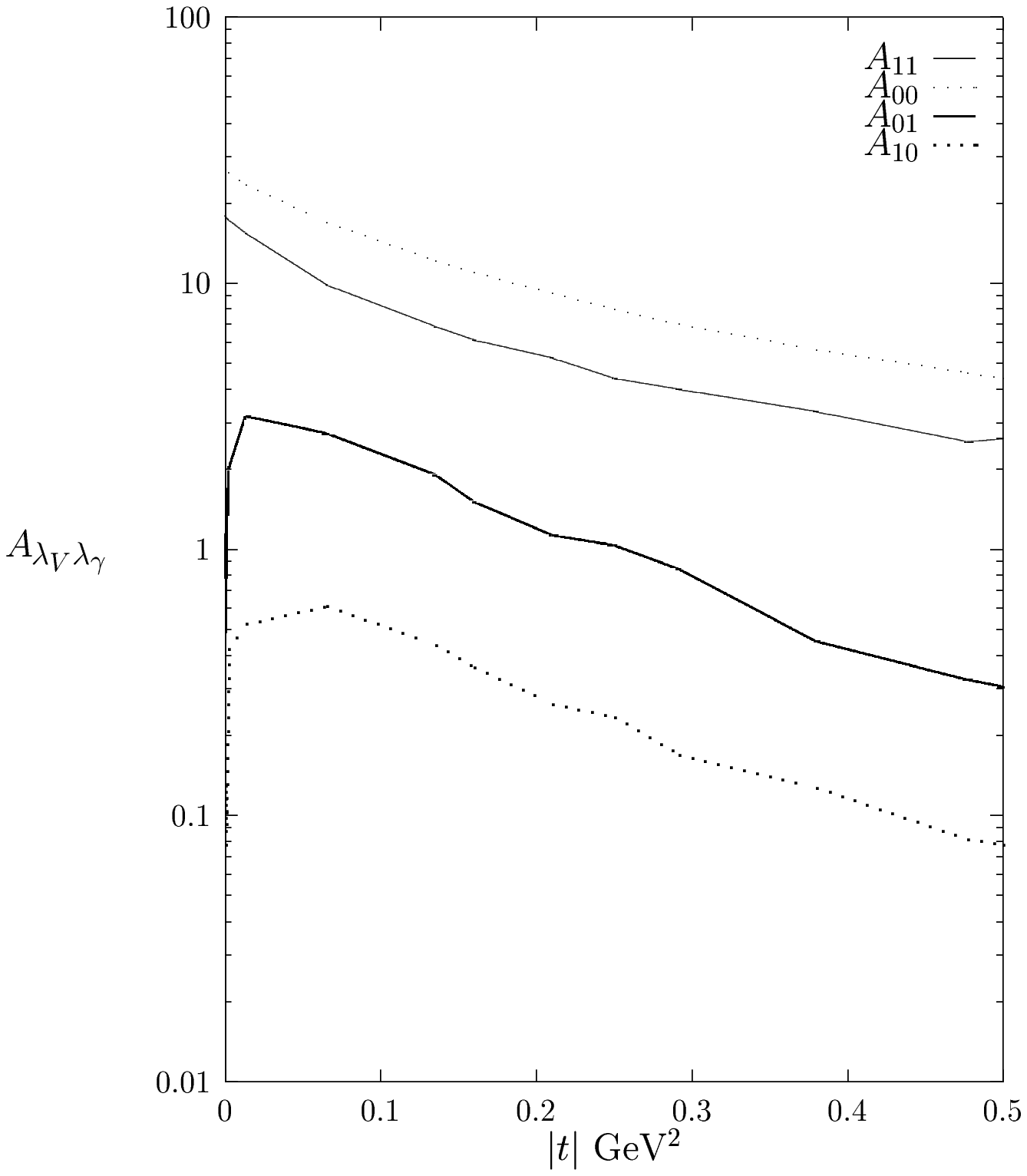,height=8cm}}
\salut{$|t|$ dependence of the helicity amplitudes for $\langle Q^2 \rangle=4.8$ GeV$^2$. \label{fig:T_t}} 
\end{figure}

\noindent
Figure~\ref{fig:T_t} shows the $|t|$ dependence of all the helicity amplitudes for $Q^2$ fixed ($\langle Q^2 \rangle=4.8$ GeV$^2$).
The single spin-flip amplitudes - $A_{01}$ and $A_{10}$ - cancel when $|t|=0$, present a maximum at very small $|t| < 0.1$ GeV$^2$ before decreasing for larger value of $|t|$.
\\

\noindent
Hence, in the high-$Q^2$ limit around $|t|=0$, the ratio of $T_{01}$ (where a longitudinal meson is produced by a transverse photon) and $T_{00}$ (where a longitudinal meson is produced by a longitudinal photon), is the following:

\begin{equation}
{T_{01} \over T_{00}}={\beta \sqrt{|t|}\over \sqrt{2} (1+\beta)\ Q}.
\end{equation}

\section{Matrix elements.}
We have seen that spin-flip amplitudes are different from zero, in particular the amplitude $A_{01}$, where a longitudinal meson is produced by a transverse photon.  Those amplitudes play a significant role in the behaviour of the 15 matrix elements measured by HERA.\\

\noindent
Following \cite{shillings} and the relations presented in the appendix~\ref{sec:appendhel}, we can derive the independent matrix elements $r_{ik}^\alpha$.\\

\noindent
Table \ref{table:helicite} present the results obtained with our model (where $A_{-11}=A_{1-1}$ are neglected) for the different matrix elements in comparison with experimental data and the SCHC assumption in the case of $\rho$ meson production.\\

\begin{figure}[H]
\begin{eqnarray}
\psfig{figure=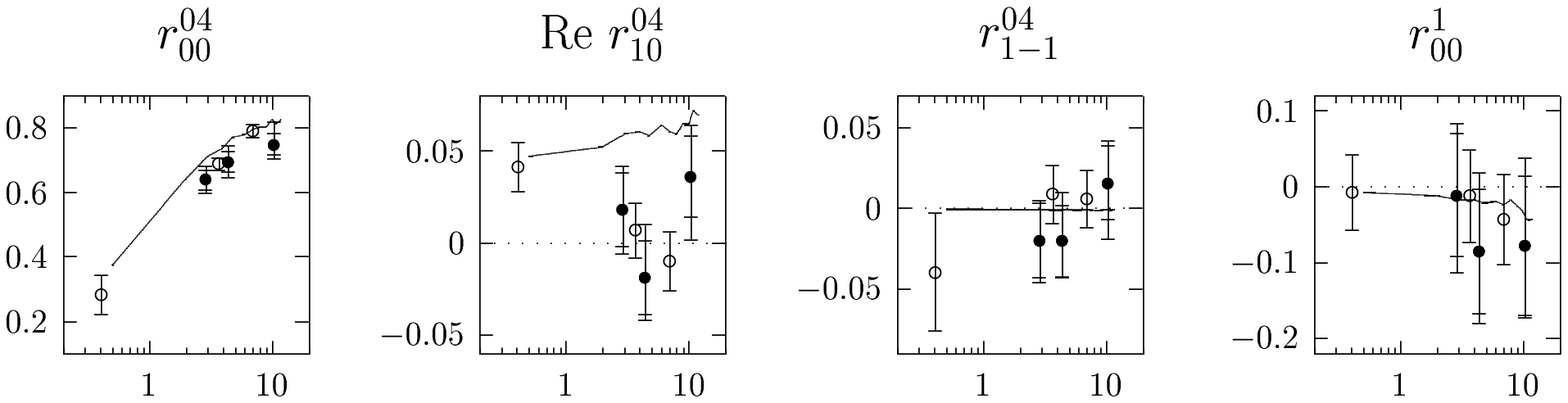,height=3.5cm}\nonumber\\
\nonumber\\
\psfig{figure=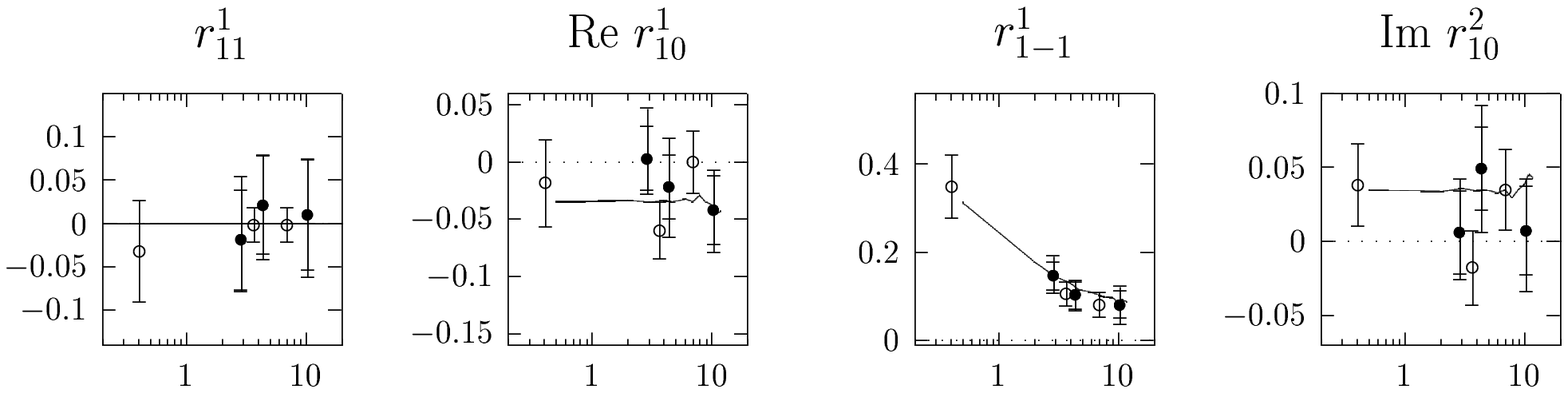,height=3.5cm}\nonumber\\
\nonumber\\
\psfig{figure=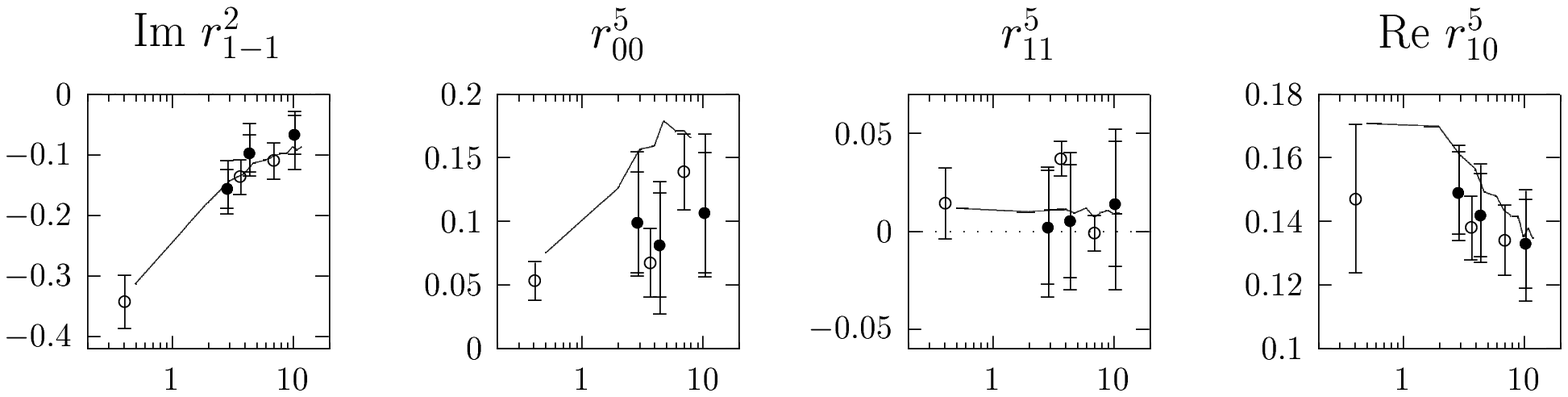,height=3.5cm}\nonumber\\
\nonumber\\
\psfig{figure=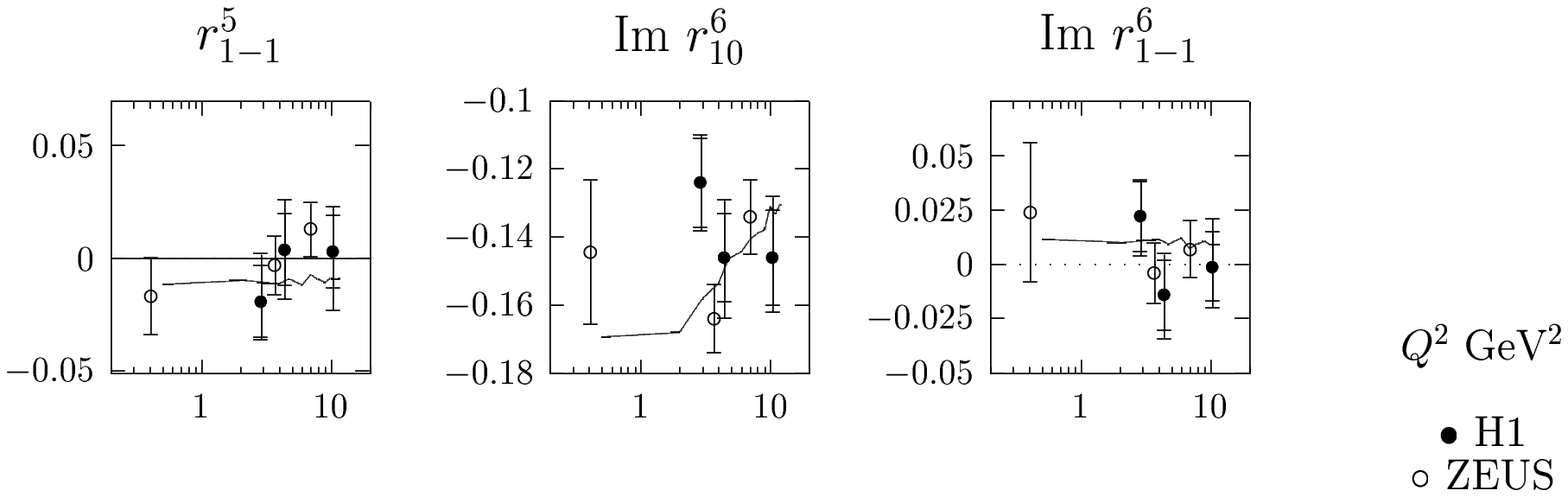,height=4.cm}\nonumber\\
\nonumber
\end{eqnarray}
\salut{The $Q^2$ dependence of the 15 spin density matrix elements for the {\it quasi-}elastic $\rho$ meson production compared to H1 \cite{Barbara} and ZEUS \cite{helZEUS} data.  The dashed line indicate the SCHC assumption.  The solid lines are the predictions obtained with our model for $|t|=0.138$ GeV$^{2}$. \label{fig:matricesQ2}}
\end{figure}

\begin{figure}[H]
\begin{eqnarray}
\psfig{figure=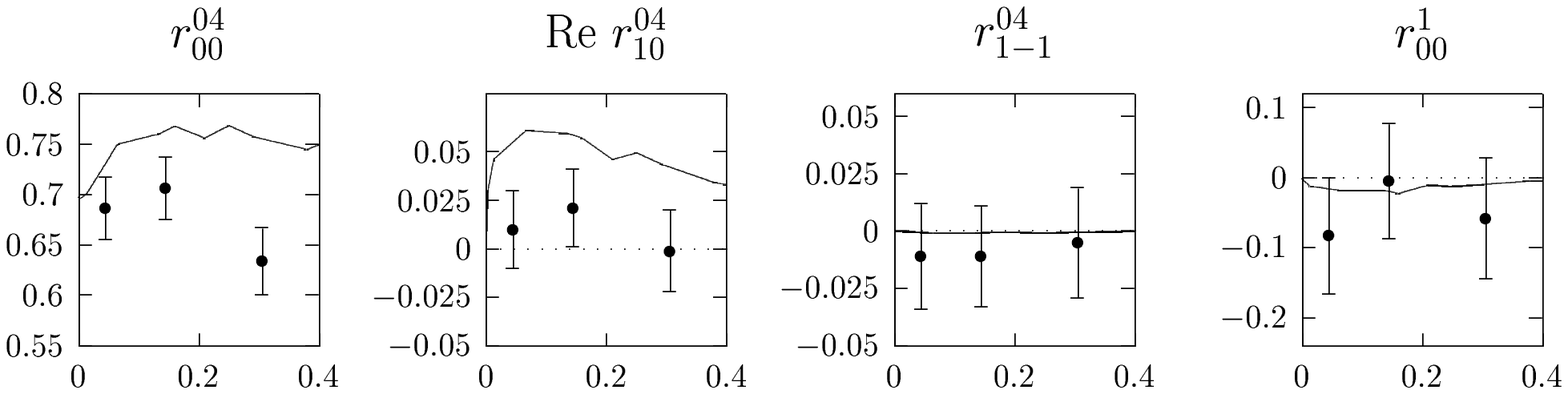,height=3.5cm}\nonumber\\
\nonumber\\
\psfig{figure=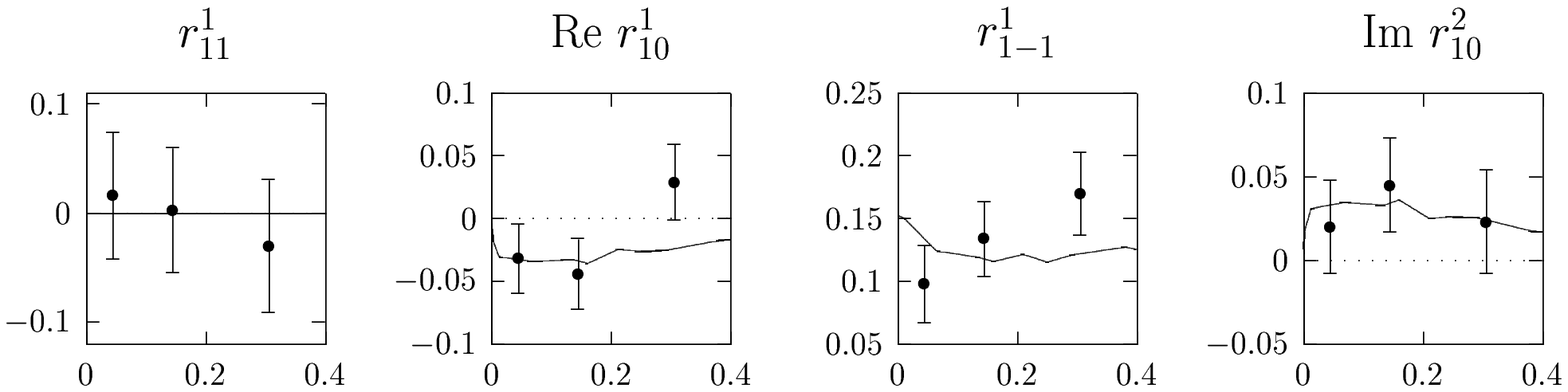,height=3.5cm}\nonumber\\
\nonumber\\
\psfig{figure=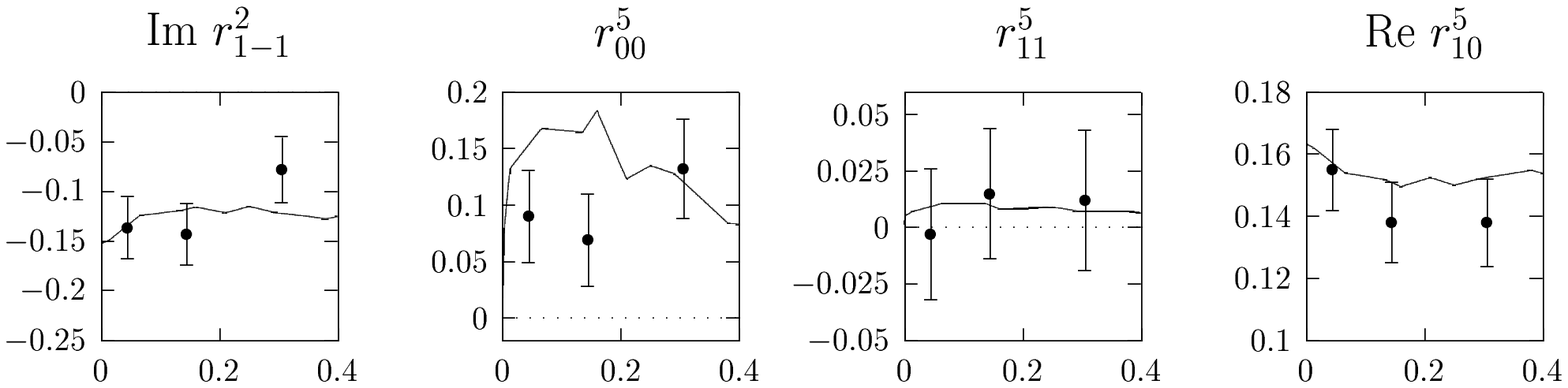,height=3.5cm}\nonumber\\
\nonumber\\
\psfig{figure=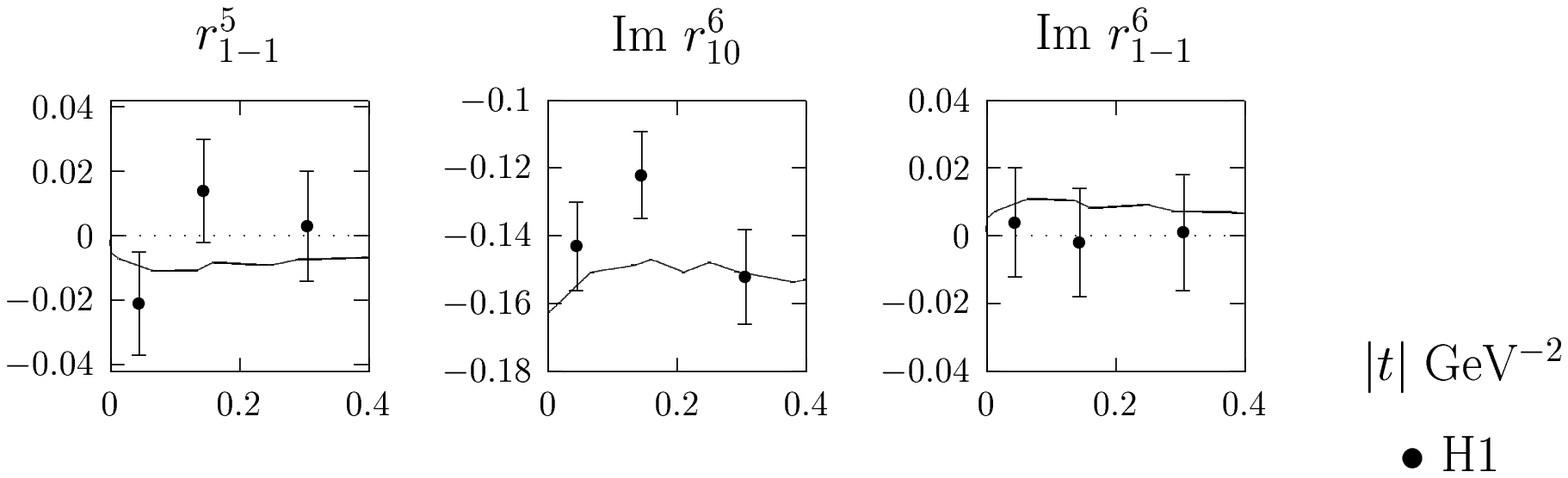,height=3.5cm}\nonumber\\
\nonumber
\end{eqnarray}
\salut{The $|t|$ dependence of the 15 spin density matrix elements for the {\it quasi-}elastic $\rho$ meson production compared to H1 \cite{Barbara} data.  The dashed line indicate the SCHC assumption.  The solid lines are the predictions obtained with our model for $Q^2=4.8$ GeV$^2$. \label{fig:matricest}}
\end{figure}

\noindent  
Figures~\ref{fig:matricesQ2} and \ref{fig:matricest} illustrate respectively the $Q^2$ and $|t|$ dependence of the 15 spin density matrix elements in the case of $\rho$ meson production.  
We can see that the model is in good agreement with the data from H1 \cite{Barbara} and ZEUS \cite{helZEUS} even at low $Q^2$.  The presence of an SCHC violation is clearly observed for $r_{00}^5$ which is significantly different from zero.  This element is indeed proportional to the most important spin-flip amplitude $A_{01}$.\\

\begin{figure}[H]
\begin{eqnarray}
\psfig{figure=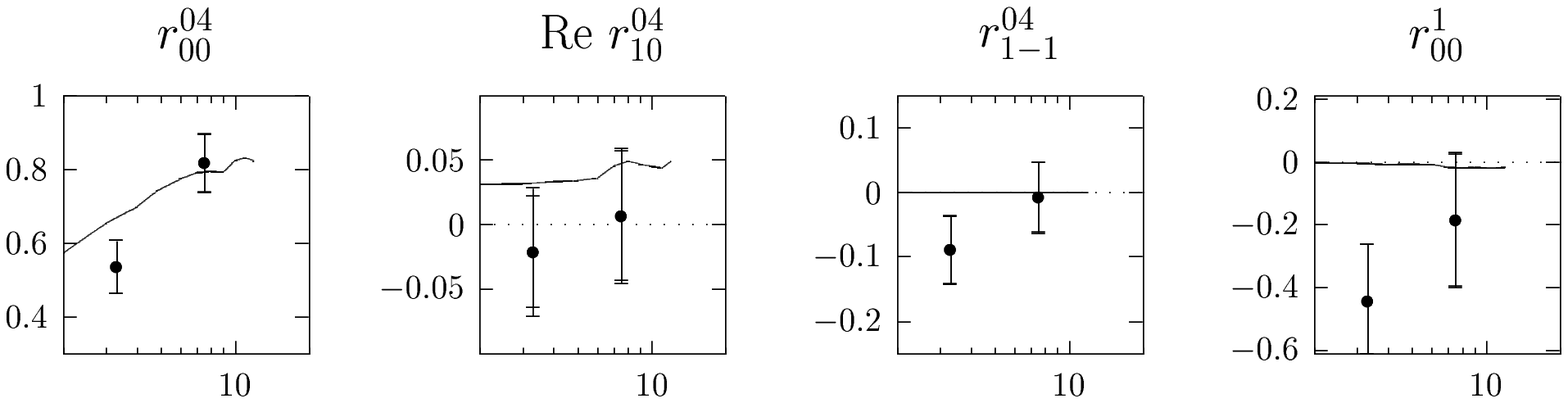,height=3.5cm}\nonumber\\
\nonumber\\
\psfig{figure=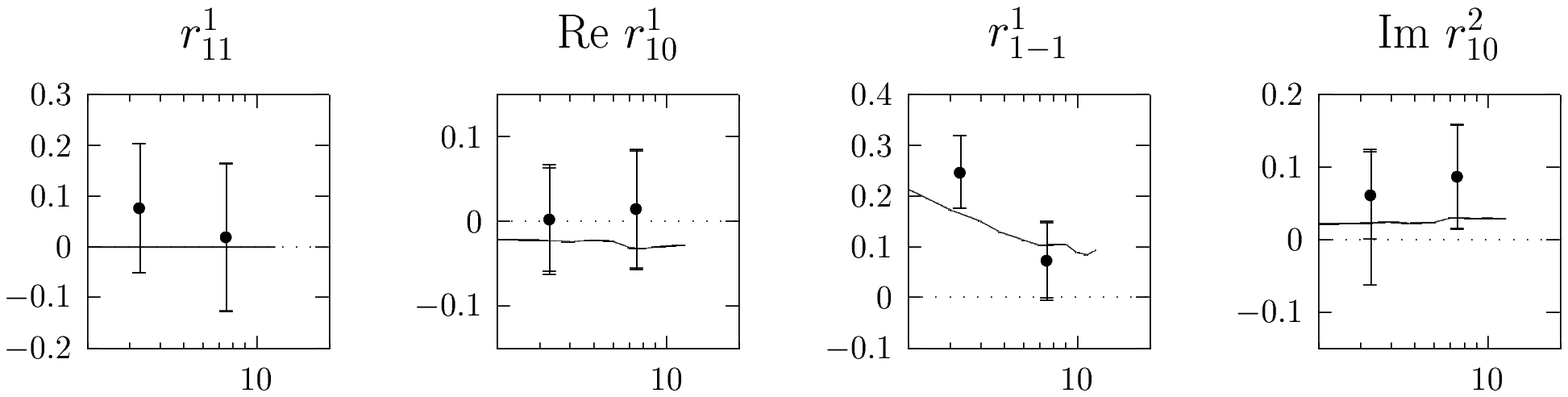,height=3.5cm}\nonumber\\
\nonumber\\
\psfig{figure=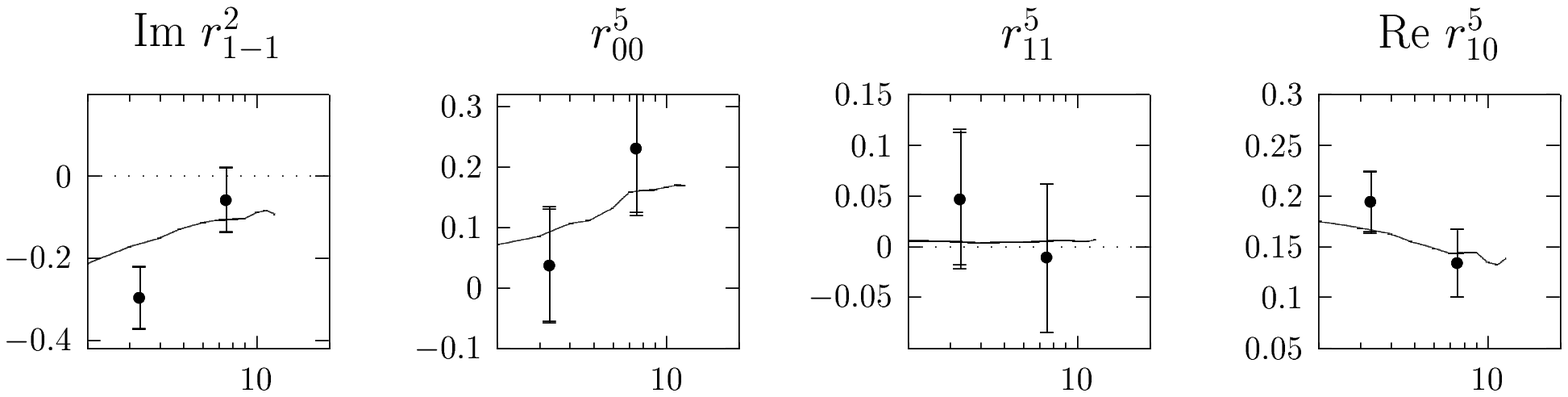,height=3.5cm}\nonumber\\
\nonumber\\
\psfig{figure=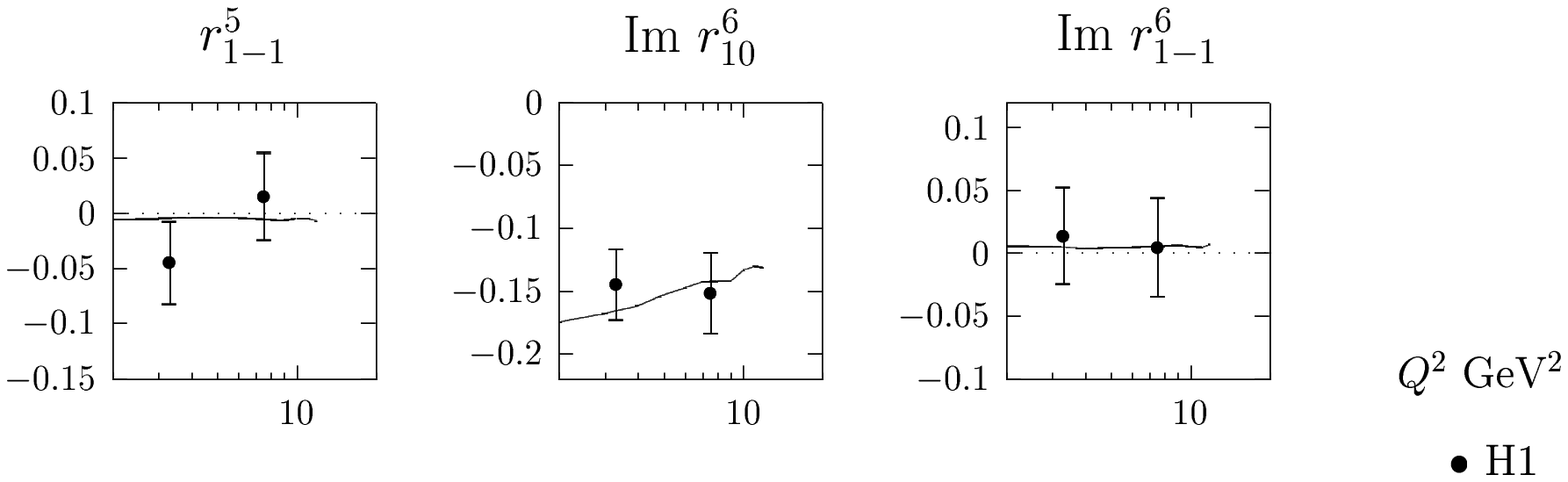,height=4.cm}\nonumber\\
\nonumber
\end{eqnarray}
\salut{The $Q^2$ dependence of the 15 spin density matrix elements for the {\it quasi-}elastic $\phi$ meson production compared to H1 \cite{H1new} data.  The dashed line indicate the SCHC assumption.  The solid lines are the predictions obtained with our model for $|t|=0.138$ GeV$^{2}$. \label{fig:matQ2}}
\end{figure}

\noindent
We made also some prediction for the production of the vector meson $\phi$.  Figure~\ref{fig:matQ2} illustrates the $Q^2$ dependence of the 15 spin density matrix elements in the case of $\phi$ meson production.

\section{Conclusion.}
Although helicity conserving amplitudes $A_{00}$, $A_{11}$ are dominant, we showed that we cannot neglect the spin-flip amplitudes.  Among them, the single spin-flip amplitude $A_{01}$ where a longitudinal meson is produced by a transverse photon is the most important.  Its presence is the indication of an evident violation of the SCHC assumption in the case of $\rho^0$ meson production.\\

\noindent
We also estimated the violation taking the ratio of the dominant spin-flip amplitude and the helicity conserving amplitudes:
\begin{eqnarray}
r={|A_{01}|  \over \sqrt{|A_{11}|^2 + |A_{00}|^2}} \approx r_{00}^5 \sqrt{{1 +R \over 2 R}}&\approx& 8.0 \pm 3.0 \% \ \ {\rm (exp)}\\
&\approx& 14. \pm 0.8 \%\ \ {\rm (model)}
\end{eqnarray}
calculated for $\langle Q^2 \rangle=4.8$ GeV$^2$ and $\langle |t| \rangle=0.138$ GeV$^2$.

\noindent
The calculation of the different helicity amplitudes and the different spin density matrix elements allows us to compare our model with H1 and ZEUS observation, and to see, once again, a good agreement.

\section*{Acknowledgments}
I thank J.R. Cudell for several useful discussions, and Barbara Clerbaux for providing me with her analyses of the data.
\begin{table}[H]
\begin{center}
\begin{tabular}{|c||c|ccc|c|c|}\hline
\multicolumn{2}{|c|}{\rule[-4mm]{0mm}{11mm}Elements}&\multicolumn{3}{|c|}{data}&our model& SCHC et Parity \\ \hline\hline
{\rule[-4mm]{0mm}{11mm}1}&$r_{00}^{04}$&0.674&$\pm$ 0.018&${-0.036 \atop +0.051}$&0.76&\large ${\epsilon R \over 1+\epsilon R}$\\ \hline
{\rule[-4mm]{0mm}{11mm}2}&Re $r_{10}^{04}$&0.011&$\pm$ 0.012&$_{-0.001}^{+0.007}$&0.059&$0$\\ \hline
{\rule[-4mm]{0mm}{11mm}3}&$r_{1-1}^{04}$&-0.010&$\pm$ 0.013&$_{-0.003}^{+0.004}$&-0.001&0\\ \hline
{\rule[-4mm]{0mm}{11mm}4}&$r_{00}^{1}$&-0.058&$\pm$ 0.048&$_{-0.011}^{+0.013}$&-0.018&0\\ \hline
{\rule[-4mm]{0mm}{11mm}5}&$r_{11}^{1}$&0.002&$\pm$ 0.034&$_{-0.006}^{+0.006}$&0.&0\\ \hline
{\rule[-4mm]{0mm}{11mm}6}&Re $r_{10}^{1}$&-0.018&$\pm$ 0.016&$_{-0.014}^{+0.010}$&-0.033& 0\\ \hline
{\rule[-4mm]{0mm}{11mm}7}&$r_{1-1}^{1}$&0.122&$\pm$ 0.018&$_{-0.005}^{+0.004}$&0.119&\large ${1 \over 2}{1 \over 1+\epsilon R}$\\ \hline
{\rule[-4mm]{0mm}{11mm}8}&Im $r_{10}^{2}$&0.023&$\pm$ 0.016&$_{-0.009}^{+0.010}$&0.033&0\\ \hline
{\rule[-4mm]{0mm}{11mm}9}&Im $r_{1-1}^{2}$&\ -0.119&$\pm$ 0.018&$_{-0.005}^{+0.010}$\ &-0.119&$-r_{1-1}^1$\\ \hline
{\rule[-4mm]{0mm}{11mm}10}&$r_{00}^{5}$&0.093&$\pm$ 0.024&$_{-0.010}^{+0.019}$&0.16&0\\ \hline
{\rule[-4mm]{0mm}{11mm}11}&$r_{11}^{5}$&0.008&$\pm$ 0.017&$_{-0.012}^{+0.008}$&0.01&0\\ \hline
{\rule[-4mm]{0mm}{11mm}12}&Re $r_{10}^{5}$&0.146&$\pm$ 0.008&$_{-0.006}^{+0.006}$&0.15&\ \large ${\sqrt{2} \over 4}{\sqrt{R}\over 1 +\epsilon R}{Re (A_{11}A_{00}^\dagger) \over |A_{11}||A_{00}|}$\\ \hline
{\rule[-4mm]{0mm}{11mm}13}&$r_{1-1}^{5}$&-0.004&$\pm$ 0.009&$_{-0.003}^{+0.001}$&-0.01&0\\ \hline
{\rule[-4mm]{0mm}{11mm}14}&Im $r_{10}^{6}$&-0.140&$\pm$ 0.008&$_{-0.004}^{+0.002}$&-0.149&-Re $r_{10}^5$\\ \hline
{\rule[-4mm]{0mm}{11mm}15}&\ Im $r_{1-1}^{6}\ $&\ 0.002&$\pm$ 0.009 &$_{-0.000}^{+0.003}$&0.01&0\\ \hline
\end{tabular}
\salut{The 15 density spin matrix elements for the elastic $\rho$ meson production for $Q^2=4.8$ GeV$^2$ and $|t|=0.138$ GeV$^2$ at H1 \cite{Barbara} in comparison with our predictions (where $A_{-11}$ and $A_{1-1}$ are neglected) and the SCHC assumption and parity conservation. \label{table:helicite}}
\end{center}
\end{table}

\appendix

\section{Relation between the 15 spin density matrix elements and the helicity amplitudes.}
\label{sec:appendhel}
The expressions of the 15 spin density matrix elements given in this appendix are derived from the appendix A of \cite{shillings}.
\\

\noindent
The normalisation factors are defined as follow:
\begin{eqnarray}
N_T&=&{1 \over 2} \sum_{\lambda_V, \lambda_{N'}, \atop \lambda_\gamma=\pm 1, \lambda_N}\ |A_{\lambda_V \lambda_{N'}, \lambda_\gamma \lambda_N}|^2\nonumber\\
&=&{1 \over 2}[\ |A_{11}|^2+|A_{-1-1}|^2+|A_{01}|^2+|A_{0-1}|^2+|A_{-11}|^2+|A_{1-1}|^2\ ]\\
N_L&=&\sum_{\lambda_V, \lambda_{N'} \lambda_N}\ |A_{\lambda_V \lambda_{N'}, 0 \lambda_N}|^2\nonumber\\
&=&|A_{00}|^2+|A_{10}|^2+|A_{-10}|^2
\end{eqnarray}
The matrix elements are combinations of helicity amplitudes:
\begin{eqnarray}
r_{00}^{04}&=&{1 \over 1+\epsilon R}[{1 \over 2 N_T}(|A_{01}|^2+|A_{0-1}|^2)+{\epsilon R \over N_L}|A_{00}|^2]  \\
Re\ r_{10}^{04}&=&{1 \over 1+\epsilon R}\ Re\ [{1 \over 2 N_T}(A_{11}A_{01}^\dagger + A_{1-1}A_{0-1}^\dagger) +{\epsilon R \over N_L} A_{10}A_{00}^\dagger] \\
r_{1-1}^{04}&=&{1 \over 1+\epsilon R}\ Re\ [{1 \over 2 N_T}(A_{11}A_{-11}^\dagger + A_{1-1}A_{-1-1}^\dagger) +{\epsilon R \over N_L} A_{10}A_{-10}^\dagger] \\
r_{00}^{1}&=&{1 \over 1+\epsilon R}\ {1 \over 2 N_T}(A_{0-1}A_{01}^\dagger + A_{01}A_{0-1}^\dagger) \\
r_{11}^{1}&=&{1 \over 1+\epsilon R}\ {1 \over 2 N_T}(A_{1-1}A_{11}^\dagger + A_{11}A_{1-1}^\dagger) \\
Re\ r_{10}^{1}&=&{1 \over 1+\epsilon R}\ {1 \over 2 N_T}\ Re\ (A_{1-1}A_{01}^\dagger + A_{11}A_{0-1}^\dagger) \\
r_{1-1}^{1}&=&{1 \over 1+\epsilon R}\ {1 \over 2 N_T}(A_{1-1}A_{-11}^\dagger + A_{11}A_{-1-1}^\dagger) \\
Im\ r_{10}^{2}&=&{1 \over 1+\epsilon R}\ {1 \over 2 N_T}\ Im\ [i(A_{1-1}A_{01}^\dagger - A_{11}A_{0-1}^\dagger)] \\
Im\ r_{1-1}^{2}&=&{1 \over 1+\epsilon R}{1 \over 2 N_T}Im [i(A_{1-1}A_{-11}^\dagger - A_{11}A_{-1-1}^\dagger)] \\
r_{00}^{5}&=&{\sqrt{R} \over 1+\epsilon R}{1 \over \sqrt{2 N_T N_L}}[Re (A_{00}A_{01}^\dagger -Re (A_{00}A_{0-1}^\dagger)] \\
r_{11}^{5}&=&{\sqrt{R} \over 1+\epsilon R}{1 \over \sqrt{2 N_T N_L}}[Re (A_{10}A_{11}^\dagger -Re (A_{10}A_{1-1}^\dagger) \\
Re\ r_{10}^{5}&=&{\sqrt{R} \over 1+\epsilon R}{1 \over \sqrt{2 N_T N_L}}{1 \over 2} Re (A_{10}A_{01}^\dagger + A_{11}A_{00}^\dagger - A_{10}A_{0-1}^\dagger - A_{1-1}A_{00}^\dagger)\nonumber \\
\\
r_{1-1}^{5}&=&{\sqrt{R} \over 1+\epsilon R}{1 \over \sqrt{2 N_T N_L}}{1 \over 2} (A_{10}A_{-11}^\dagger + A_{11}A_{-10}^\dagger - A_{10}A_{-1-1}^\dagger - A_{1-1}A_{-10}^\dagger)\nonumber\\
 \\
Im\ r_{10}^{6}&=&{\sqrt{R} \over 1+\epsilon R}{1 \over \sqrt{2 N_T N_L}}{1 \over 2} Re (A_{10}A_{01}^\dagger - A_{11}A_{00}^\dagger + A_{10}A_{0-1}^\dagger - A_{1-1}A_{00}^\dagger)\nonumber \\
\\
Im\ r_{1-1}^{6}&=&{\sqrt{R} \over 1+\epsilon R}{1 \over \sqrt{2 N_T N_L}}{1 \over 2} Re (A_{10}A_{-11}^\dagger - A_{11}A_{-10}^\dagger + A_{10}A_{-1-1}^\dagger - A_{1-1}A_{-10}^\dagger)\nonumber \\
\end{eqnarray}
where $R$ is the ratio of the longitudinal and transverse cross sections $\gamma^* p$:
\begin{equation}
R={N_L \over N_T}={\sigma_L \over \sigma_T}.
\end{equation}
\end{document}